\documentclass{article}
\usepackage{amsmath,amssymb,enumerate,epsfig,bbm}

\newcommand{\TeXmacs}{T\kern-.1667em\lower.5ex\hbox{E}\kern-.125emX\kern-.1em\lower.5ex\hbox{\textsc{m\kern-.05ema\kern-.125emc\kern-.05ems}}}
\newcommand{\email}[1]{{\textit{Email:} \texttt{#1}}}
\newcommand{\keywords}[1]{{\textbf{Keywords:} #1}}
\newcommand{\mathe}{\mathrm{e}}
\newcommand{\tmem}[1]{{\em #1\/}}
\newcommand{\tmmathbf}[1]{\ensuremath{\boldsymbol{#1}}}
\newcommand{\tmop}[1]{\ensuremath{\operatorname{#1}}}
\newcommand{\tmstrong}[1]{\textbf{#1}}

\newcommand{\withTeXmacstext}{This document has been produced using \TeXmacs (see \texttt{http://www.texmacs.org})}
\newenvironment{enumerateroman}{\begin{enumerate}[i.] }{\end{enumerate}}
\newenvironment{proof}{\noindent\textbf{Proof\ }}{\hspace*{\fill}$\Box$\medskip}
\newtheorem{lemma}{Lemma}
\newtheorem{theorem}{Theorem}

\DeclareGraphicsExtensions{.eps}

\begin{document}

\title{On the Kert\'esz line: Some rigorous bounds\thanks{{\withTeXmacstext}}
\thanks{\keywords{Ising model, Potts model, percolation, random cluster model, random
media, phase transition.}}}
\author{Jean RUIZ\thanks{\email{ruiz@cpt.univ-mrs.fr}}\\
Centre de Physique Th\'eorique, CNRS Luminy case 907,\\ F--13288 Marseille Cedex
9, France.\and Marc WOUTS\thanks{\email{marc.wouts@u-paris10.fr}}\\
Modal'X, Universit\'e Paris Ouest - Nanterre La D\'efense. B\^at. G. \\
200 avenue de la R\'epublique, 92001 Nanterre Cedex.}\maketitle

\begin{abstract}
  We study the Kert\'esz line of the $q$--state Potts model at (inverse)
  temperature $\beta$, in presence of an external magnetic field $h$. This line
  separates two regions of the phase diagram according to the existence or not
  of an infinite cluster in the Fortuin-Kasteleyn representation of the model.
  It is known that the Kert\'esz line $\text{$h_K (\beta)$}$ coincides with
  the line of first order phase transition for small fields when $q$ is large
  enough. Here we prove that the first order phase transition implies a
  {\tmem{jump}} in the density of the infinite cluster, hence the Kert\'esz
  line remains below the line of first order phase transition. We also analyze
  the region of large fields and prove, using techniques of stochastic
  comparisons, that $h_K (\beta)$ equals $\log (q - 1) - \log (\beta -
  \beta_p)$ to the leading order, as $\beta$ goes to $\beta_p = - \log (1 -
  p_c)$ where $p_c$ is the threshold for bond percolation.
\end{abstract}

One important feature of the Fortuin--Kasteleyn representation of Ising and
Potts models {\cite{FK1}} (the random cluster model), is that the geometrical
transition, i.e. the apparition of an infinite cluster, corresponds precisely
to the phase transition leading to a spontaneous magnetization in the absence
of an external field {\cite{CK}}. In {\cite{Kertesz1}}, Kert\'esz pointed out
that this property is lost in the Ising model when an external field $h$ is
introduced: while thermodynamic quantities are analytic for any $h > 0$, a
geometric transition appears in the corresponding random cluster model and
there is a whole percolation transition line extending from the Curie point
($h = 0$) to infinite fields. As Kert\'esz explained, the analyticity of
thermodynamic quantities and the existence of the percolation transition are
not contradictory because the free energy remains analytic.

The Kert\'esz line can be considered as well in the Potts model. There, for
large $q$, the first order transition extends to small, positive fields $h$
and it is an important issue to understand whether or not the Kert\'esz line
coincides with the line of phase transition. Such a property was established
in {\cite{BGLRS}} for small $h$ (and $q$ large enough) and hence extends the
relevance of the random cluster representation for the analysis of the phase
transition in the corresponding region. Here we address some of the remaining
issues: we prove the existence of the line, show that the first order phase
transition results in a discontinuity of the percolation density, and provide
bounds on the Kert\'esz line that are particularly precise in the region of
large fields.

In the Potts model, the spin variables $\sigma_i$ associated with lattice
sites take values in the discrete set \{$1, \ldots ., q\}$. Considering a spin
configuration in a finite box $\Lambda \subset \mathbbm{Z}^d$ ($d \geqslant
2$), the Potts model at inverse temperature $\beta$, subject to an external
ordering field $h$, is defined by the Gibbs measure
\begin{equation}
  \mu_{\Lambda}^{\tmop{Potts}} (\tmmathbf{\sigma}) =
  \frac{1}{Z_{\Lambda}^{\tmop{Potts}}} \prod_{\langle i, j \rangle}^{_{}}
  e^{\beta (\delta_{\sigma_i, \sigma_j} - 1)} \prod_i^{_{}} e^{h
  \delta_{\sigma_i, 1}} \label{Potts}
\end{equation}
Here the first product is over nearest neighbor pairs of $\Lambda$, the second
runs over sites of $\Lambda$, $Z_{\Lambda}^{\tmop{Potts}}$ denotes the
partition function (normalizing factor) and $\delta$ is the Kronecker symbol.

To study the behavior of clusters, in the sense of FK clusters, we turn to the
corresponding Edwards--Sokal formulation {\cite{ES}}, given by the joint
measure
\begin{equation}
  \mu_{\Lambda}^{\tmop{ES}} (\tmmathbf{\sigma}, \tmmathbf{\eta}) =
  \frac{1}{Z_{\Lambda}^{\tmop{ES}}} \prod_{\langle i, j \rangle}^{_{}} \left(
  e^{- \beta} \delta_{\eta_{i j}, 0} + (1 - e^{- \beta}) \delta_{\eta_{i j},
  1} \delta_{\sigma_i, \sigma_j} \right) \prod_i^{_{}} e^{h \delta_{\sigma_i,
  1}} \label{ES} .
\end{equation}

This model can be thought as follows. Given a spin configuration, between two
neighboring sites with $\sigma_i = \sigma_j$, one put an edge ($\eta_{i j} =
1$) with probability $1 - e^{- \beta}$ and no edge w.p. $e^{- \beta}$; for
$\sigma_i \neq \sigma_j$, no edge (or bond) is present. When the field is
infinite all spins take the value one and we are left with the classical bond
percolation problem. At finite fields, the spins are not uniformly equal to
one yet we will see that percolation in the edge variable $\tmmathbf{\eta}$
still occurs at some finite temperature.

We call $\mu_{\Lambda, \tmmathbf{f}}^{\tmop{RC}}$ the marginal law of
$\tmmathbf{\eta}$ under $\mu_{\Lambda}^{\tmop{ES}}$. This measure can be
considered as well for non-integer $q \geqslant 1$ (see (\ref{RC})). Our first
result concerns the existence of the Kert\'esz line.

\begin{theorem}
  \label{thm-ex}Let $\beta, h \geqslant 0$ and $q \geqslant 1$.
  \begin{enumerateroman}
    \item The infinite volume limit
    \begin{equation}
      \mu_{\tmmathbf{f}}^{\tmop{RC}} = \lim_{\Lambda \nearrow \mathbbm{Z}^d}
      \mu_{\Lambda, \tmmathbf{f}}^{\tmop{RC}} \label{muf}
    \end{equation}
    exists.
    
    \item The probability
    \begin{equation}
      \theta = \mu_{\tmmathbf{f}}^{\tmop{RC}} \left( \text{the origin belongs
      to an infinite cluster of } \tmmathbf{\eta} \right) \label{theta}
    \end{equation}
    increases with $\beta$ and $h$, and decreases with $q$.
    
    \item Hence the Kert\'esz line
    \begin{equation}
      h_K (\beta) = \inf \{h \geqslant 0 : \theta > 0\}
    \end{equation}
    exists, and $h_K (\beta)$ decreases with $\beta$.
  \end{enumerateroman}
\end{theorem}

Note that $h_K (\beta) = 0$ if $\beta \geqslant \beta_c$, where $\beta_c$ is
the critical inverse temperature for the phase transition with no field, while
$h_K (\beta) = + \infty$ if $\beta \leqslant \beta_p$, where
\begin{equation}
  \beta_p = - \log (1 - p_c)
\end{equation}
is the critical inverse temperature for percolation at infinite fields and
$p_c$ the threshold for bond percolation on $\mathbbm{Z}^d$.

Then we examine the consequences of the first order transition on the density
$\theta$ of the infinite cluster:

\begin{theorem}
  \label{thm-first-order}A discontinuity in the parameter $\beta$ in the mean
  energy
  \begin{equation}
    e_{\tmmathbf{f}} = \frac{1}{1 - e^{- \beta}}
    \mu_{\tmmathbf{f}}^{\tmop{RC}} \left( \eta_{i j} \right) \label{ef}
  \end{equation}
  where $i, j$ are neighboring sites, or in the magnetization, implies a
  discontinuity in the density $\theta$ of percolation.
\end{theorem}

This means that $\theta$ has a {\tmem{jump}} on the line of first order phase
transition. Consequently, the Kert\'esz line cannot be found above the line of
first order phase transition. It is known {\cite{BGLRS}} that both lines
coincide at small fields when $q$ is large, hence the question remains whether
they coincide up to the other extremity of the line of first order phase
transition. In the corresponding mean field analysis {\cite{BGRW}} we proved
the existence of a cusp as soon as $q > 2$, that is, whenever appears a line
of first order phase transition. However, in the two dimensional Potts model
no bifurcation was noted numerically {\cite{BGLRS}}.

We conclude our exposition of the results with upper and lower bounds on the
Kert\'esz line, which are particularly efficient when $\beta$ is taken
slightly above $\beta_p = - \log (1 - p_c)$, corresponding to the regime of
large fields. The idea that led to the next theorem is that the model can be
understood as independent bond percolation over a {\tmem{random media}} : the
spins not equal to $1$ are considered as {\tmem{defects}}, which become rare
when $h \rightarrow + \infty$. Our proofs are reminiscent of {\cite{ACCN}} in
which similar methods were employed to provide necessary and sufficient
conditions for the phase transition in the dilute Ising model, see also
{\cite{CS}} for beautiful results on mixed percolation.

\begin{theorem}
  For any $d \geqslant 2$, $q > 1$ and $\beta > \beta_p$, one has
  \begin{eqnarray}
    h_K (\beta) & \leqslant & - \log \frac{\sqrt{\frac{e^{\beta} -
    1}{e^{\beta_p} - 1}} - 1}{q - 1}  \label{eq-upb-BK}\\
    & = & - \log \left( \beta - \beta_p \right) + \log \left( 2 p_c (q - 1)
    \right) + O_{\beta \rightarrow \beta_p^+} (\beta - \beta_p) \\
    \text{while \ \ \ \ \ } h_K (\beta) & \geqslant & - \log \frac{e^{-
    \beta_p} - \mathe^{- \beta}}{p_c (q - 1)} - 2 \beta d  \label{eq-lwb-BK}\\
    & = & - \log \left( \beta - \beta_p \right) + \log \left( p_c (q - 1)
    \right) - (2 d - 1) \beta_p + O_{\beta \rightarrow \beta_p^+} (\beta -
    \beta_p) . 
  \end{eqnarray}
\end{theorem}

Thus, to the leading order, $h_K (\beta) \simeq - \log (\beta - \beta_p) +
\log (q - 1)$ when $\beta \rightarrow \beta_p^+$. The upper and lower
asymptotes differ from the constant $\log (2) - (2 d - 1) \log (1 - p_c)$ that
does not depend on $q$.

These upper and lower bounds are presented in Fig. 1 together with the
numerical results of {\cite{BGLRS}}.

\begin{figure}[h]
\begin{center}
  \includegraphics[width=8cm]{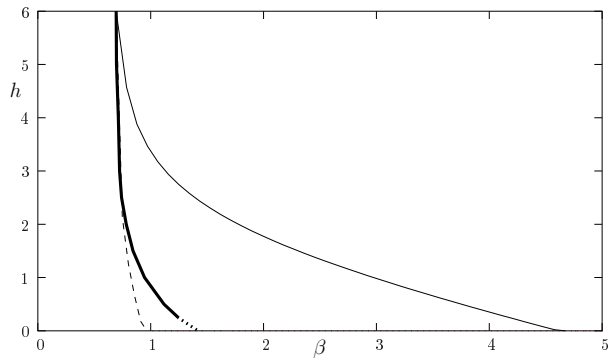}
  \caption{A comparison between upper and lower bounds with the numerical
  results of {\cite{BGLRS}} for $d = 2$ and $q = 10$.}
\end{center}
\end{figure}

To summarize, we have shown that for the lattice Potts model subject to an
external field, the Kert\'esz line is well defined. We have presented upper and
lower bounds on this line. These bounds are very precise at high fields and
complement the previous study {\cite{BGLRS}} in which a precise approximation
at low field was given. In addition, we have shown that a jump of the mean
energy or of the magnetization implies a jump in the percolation density of
the clusters associated to the corresponding FK representation of the model. \
This last result does not exclude the presence of an intermediate regime of
the field where when decreasing the temperature, one first encounters the
percolation transition and then for a lower temperature the percolation
density would exhibits a jump.

{\tmstrong{Acknowledgments.}} It is a pleasure to thank Daniel Gandolfo for
valuable discussions. One of us (M. W.) is grateful to CPT and LATP for their
kind hospitality.

\appendix\section{Appendix}

\subsection{A random cluster representation}

For any $\beta, h \geqslant 0$ and $q \geqslant 1$ we define a random cluster
model that takes into account the external field. Edge configurations
$\tmmathbf{\eta}$ have $\eta_{i j} \in \{0, 1\}$ for all $< i, j >$ nearest
neighbor pairs in the domain. For $\Lambda$ a finite subset of $\mathbbm{Z}^d$
and $\tmmathbf{\pi}$ a boundary condition on $\Lambda$, that is an edge
configuration on $\mathbbm{Z}^d$ which restriction to $\Lambda$ has no open
edge, we consider
\begin{equation}
  \mu_{\Lambda, \tmmathbf{\pi}}^{\tmop{RC}} (\tmmathbf{\eta}) =
  \frac{1}{Z_{\Lambda, \tmmathbf{\pi}}^{\tmop{RC}}} \prod_{\langle i, j
  \rangle}^{_{}} \left( e^{- \beta} \delta_{\eta_{i j}, 0} + (1 - e^{- \beta})
  \delta_{\eta_{i j}, 1}  \right) \prod_{C \in
  \mathcal{C}_{\Lambda}^{\tmmathbf{\pi}} (\tmmathbf{\eta})} w (S (C))
  \label{RC}
\end{equation}
where
\[ w (S) = 1 + (q - 1) e^{- hS} . \label{w} \]
The first product runs over $\left\langle i, j \right\rangle$ nearest neighbor
pairs in $\Lambda$. The second one is over all connected components (clusters)
$C \in \mathcal{C}_{\Lambda}^{\tmmathbf{\pi}} (\tmmathbf{\eta})$, where
$\mathcal{C}_{\Lambda}^{\tmmathbf{\pi}} (\tmmathbf{\eta})$ is the set of
clusters of $\mathbbm{Z}^d$ under the wiring $\tmmathbf{\pi} \vee
\tmmathbf{\eta}$ (the edge configuration defined by $(\pi \vee \eta)_{i j} =
\max (\pi_{i j}, \eta_{i j})$), that own some site of $\Lambda$. For any such
cluster, $S (C)$ stands for its number of sites.

The variable $\tmmathbf{\eta}$ under the joint measure
$\mu_{\Lambda}^{\tmop{ES}}$ defined at (\ref{ES}) follows the law
$\mu_{\Lambda, \tmmathbf{f}}^{\tmop{RC}}$, where $\tmmathbf{f}$ stands for the
{\tmem{free boundary condition}}, that is the edge configuration with no edge
open. Conditionally on $\tmmathbf{\eta}$, for integer $q \in \{2, 3, \ldots\}$
the distribution of $\tmmathbf{\sigma}$ under the joint measure
$\mu_{\Lambda}^{\tmop{ES}}$ is as follows: the spin is constant on each
cluster of $\tmmathbf{\eta}$, and a cluster with $S$ sites obtains the color
$1$ with probability $e^{hS} / (e^{hS} + q - 1)$, any other of the $q - 1$
colors with probability $1 / (e^{hS} + q - 1)$, independently of other
clusters. This conditional distribution accounts for the definition (\ref{mf})
of the magnetization.

Let us now compare our representation with that of {\cite{BGLRS}}. There, as
an alternative to the ghost spin scheme, a colored version of the
Edwards--Sokal representation was introduced. This representation was defined
by
\begin{eqnarray}
  \mu_{\Lambda}^{\tmop{CES}} (\tmmathbf{\sigma}, \tmmathbf{n}) & = &
  \frac{1}{Z_{\Lambda}^{\tmop{CES}}} \prod_{\langle i, j \rangle}^{_{}} [e^{-
  \beta} \delta_{n_{i j}, 0} + (1 - e^{- \beta}) \delta_{n_{i j}, 1} \chi
  (\sigma_i = \sigma_j = 1) \nonumber\\
  &  & + (1 - e^{- \beta}) \delta_{\eta_{i j}, 2} \chi (\sigma_i = \sigma_j
  \neq 1)] \prod_i^{_{}} e^{h \delta_{\sigma_i, 1}}  \label{CES}
\end{eqnarray}
with the edges variables $n_{ij}$ taking values in the set $\{0, 1, 2\}$. Let
us consider, for a while, thermodynamics limits (the existence of
thermodynamics limits will be proven at the next section). We want to
emphasize that the question of percolation for $\tmmathbf{\eta}$ under
$\mu_{\tmmathbf{f}}^{\tmop{RC}}$ and for the color $1$ in $\tmmathbf{n}$ under
$\mu^{\tmop{CES}}$ are {\tmem{equivalent}}. Indeed, any infinite cluster for
$\tmmathbf{\eta}$ under $\mu^{\tmop{ES}}$ will be given the color $\sigma = 1$
with probability one as soon as $h > 0$ (resp. w. p. $1 / q$ if $h = 0$).
Hence, relabelling $\tmmathbf{\eta}$ into $\tmmathbf{n}$ according to the spin
of clusters we obtain in fact an infinite cluster for the color $1$ in
$\tmmathbf{n}$, w. p. 1 (w. p. $1 / q$ if $h = 0$) and this shows that the
probability of percolation from the origin $\theta$ under
$\mu_{\tmmathbf{f}}^{\tmop{RC}}$ and $\theta_1$ for the color $1$ in
$\tmmathbf{n}$ under $\mu^{\tmop{CES}}$ satisfy
\[ \theta_1 = \left\{ \begin{array}{ll}
     \theta_{} / q & \text{if } h = 0\\
     \theta_{} & \text{if } h > 0.
   \end{array} \right. \]

\subsection{Conditional probabilities and infinite volume limit}

Here we give the proof of Theorem \ref{thm-ex}. Like the usual random cluster
representation, the measures $\mu_{\Lambda, \tmmathbf{\pi}}^{\tmop{RC}}$
satisfy the DLR equations, which means that, given any $\Lambda' \subset
\Lambda$, the restriction of $\tmmathbf{\eta}$ to $\Lambda'$ under the measure
$\mu_{\Lambda, \tmmathbf{\pi}}^{\tmop{RC}}$ conditioned on
$\tmmathbf{\eta}=\tmmathbf{\omega}$ outside of $\Lambda'$ has law
$\mu_{\Lambda', \tmmathbf{\omega} \vee \tmmathbf{\pi}}^{\tmop{RC}}$.
Consequently, the measures $\mu_{\Lambda, \tmmathbf{\pi}}^{\tmop{RC}}$ are
characterized by the law of $\tmmathbf{\eta}$ on a single edge $i j$ given the
boundary condition $\tmmathbf{\pi}$:
\begin{equation}
  \mu_{\{i j\}, \tmmathbf{\pi}}^{\tmop{RC}} (\tmmathbf{\eta}_{i j} = 1) = p_{i
  j}^{\tmmathbf{\pi}} \label{RC-ij}
\end{equation}
where $p_{i j}^{\tmmathbf{\pi}} = p \overset{\tmop{def} .}{=} 1 - e^{- \beta}$
if $\tmmathbf{\pi}$ connects $i$ and $j$, otherwise
\begin{equation}
  p_{i j}^{\tmmathbf{\pi}} = \frac{p}{p + (1 - p) \frac{w
  (S_i^{\tmmathbf{\pi}}) w (S_j^{\tmmathbf{\pi}})}{w (S_i^{\tmmathbf{\pi}} +
  S_j^{\tmmathbf{\pi}})}} \label{p}
\end{equation}
where $S_i^{\tmmathbf{\pi}}$ (resp. $S_j^{\tmmathbf{\pi}}$) is the number of
sites of the cluster containing $i$ (resp. $j$) under the connections
$\tmmathbf{\pi}$.

It is easily verified that $p_{i j}^{\tmmathbf{\pi}}$ is an increasing
function of $\beta$, $h$ and $\tmmathbf{\pi}$, decreasing with $q \geqslant
1$. Thanks to the DLR equations, the hypothesis of Holley's Lemma (see for
instance Theorem 4.8 in {\cite{GHM}} or Theorems 2.1 and 2.6 in {\cite{Gri}})
are verified and this implies that $\mu_{\Lambda, \tmmathbf{\pi}}^{\tmop{RC}}$
stochastically increases with $\beta, h$ and $\tmmathbf{\pi}$, and
stochastically decreases with $q$. Using again the DLR equations, we see that
the measure $\mu_{\Lambda, \tmmathbf{f}}^{\tmop{RC}}$ stochastically increases
as $\Lambda \nearrow \mathbbm{Z}^d$, proving the existence of the weak limit
$\mu_{\tmmathbf{f}}^{\tmop{RC}}$ at (\ref{muf}). Point {\tmem{ii}} of the
theorem follows from the variations of $\mu_{\tmmathbf{f}}^{\tmop{RC}}$ with
$\beta, h$ and $q$ which are the same than those of $\mu_{\Lambda,
\tmmathbf{\pi}}^{\tmop{RC}}$ while point {\tmem{iii}} is an immediate
consequence of {\tmem{ii}}.

\subsection{First order transitions}

Theorem \ref{thm-first-order} is essentially a consequence of the uniqueness
of infinite volume measures under the condition that the infinite cluster has
the same density under both infinite volume limits for free and wired boundary
conditions (Theorem \ref{thm-thetafw} below). We adapt here the classical
argument at $h = 0$ to our setting $h \geqslant 0$.

By the stochastic comparison argument, one can consider as well the infinite
volume limit $\mu_{\tmmathbf{w}}^{\tmop{RC}}$ of $\mu_{\Lambda,
\tmmathbf{\pi}}^{\tmop{RC}}$ under the wired boundary condition
$\tmmathbf{w}$, that has all edges open. As in {\cite{Leb77,Gri95,Wou07}} it
happens that:

\begin{lemma}
  \label{lem-fw}Given $h \geqslant 0, q \geqslant 1$, the set of $\beta$ at
  which $\mu^{\tmop{RC}}_{\tmmathbf{f}} \neq \mu^{\tmop{RC}}_{\tmmathbf{w}}$
  is at most countable.
\end{lemma}

\begin{proof}
  Let
  \begin{equation}
    y^{\tmmathbf{\pi}}_{\Lambda} = \frac{1}{| \Lambda |} \log \left[
    \prod_{\langle i, j \rangle}^{_{}} \left( e^{\beta} - 1 \right)^{\eta_{i
    j}} \prod_{C \in \mathcal{C}_{\Lambda}^{\tmmathbf{\pi}} (\tmmathbf{\eta})}
    w (S (C)) \right] .
  \end{equation}
  When $\tmmathbf{\pi}=\tmmathbf{f}$, the quantity
  $y^{\tmmathbf{\pi}}_{\Lambda}$ is sub-additive -- when one cluster of size
  $S$ is cut into two clusters of size $S_1, S_2$ with $S_1 + S_2 = S$, then
  $w (S) \leqslant w (S_1) w (S_2)$. Hence $y^{\tmmathbf{f}}_{\Lambda}$
  converges to some $y (\beta, h)$ as $\Lambda \rightarrow \mathbbm{Z}^d$. The
  influence of the boundary condition $\tmmathbf{\pi}$ on
  $y^{\tmmathbf{\pi}}_{\Lambda}$ is of order $| \partial \Lambda | / | \Lambda
  |$: for any configuration $\tmmathbf{\eta}$ the product $\prod_{C \in
  \mathcal{C}_{\Lambda}^{\pi} (\tmmathbf{\eta})} w (S (C))$ decreases with
  $\tmmathbf{\pi}$ and conversely,
  \begin{equation}
    \prod_{C \in \mathcal{C}_{\Lambda}^{\tmmathbf{f}} (\tmmathbf{\eta})} w (S
    (C)) \leqslant (w (1))^{| \partial \Lambda |} \prod_{C \in
    \mathcal{C}_{\Lambda}^{\tmmathbf{w}} (\tmmathbf{\eta})} w (S (C))
  \end{equation}
  because $\mathcal{C}^{\tmmathbf{f}}_{\Lambda} (\tmmathbf{\eta})$ contains at
  most $| \partial \Lambda |$ clusters not present in
  $\mathcal{C}_{\Lambda}^{\tmmathbf{w}} (\tmmathbf{\eta})$, each of them
  having size $S \geqslant 1$. Hence for any sequence
  $\tmmathbf{\pi}_{\Lambda}$, any sequence of cubes $\Lambda \rightarrow
  \mathbbm{Z}^d$, we have $y^{\tmmathbf{\pi}}_{\Lambda} \rightarrow y$. Now we
  show that $y^{\tmmathbf{\pi}}_{\Lambda}$ is a convex function of $\lambda =
  \log (e^{\beta} - 1)$. Indeed, the derivative
  \[ \frac{\partial y^{\tmmathbf{\pi}}_{\Lambda}}{\partial \lambda} =
     \mu^{\tmop{RC}}_{\Lambda, \tmmathbf{\pi}} \left( \frac{1}{| \Lambda |}
     \sum_{< i, j >} \eta_{i j} \right) \]
  is an increasing function of $\beta$, hence of $\lambda$, and the convexity
  holds for both $y^{\tmmathbf{\pi}}_{\Lambda}$ and its limit $y$. Therefore
  $y$ is derivable at all $\beta \notin \mathcal{D}_h$ where $\mathcal{D}_h$
  (that depends on $h$) is finite or countable. When this occurs, by convexity
  of the $y^{\tmmathbf{\pi}}_{\Lambda}$ we have
  \[ \lim_{\Lambda} \frac{\partial y^{\tmmathbf{f}}_{\Lambda}}{\partial
     \lambda} = \frac{\partial y_{\Lambda}}{\partial \lambda} = \lim_{\Lambda}
     \frac{\partial y^{\tmmathbf{w}}_{\Lambda}}{\partial \lambda} \]
  which implies that the probability of opening a given edge is the same under
  both free and wired boundary conditions: $\mu^{\tmop{RC}}_{\tmmathbf{f}}
  (\eta_{i j}) = \mu^{\tmop{RC}}_{\tmmathbf{w}} (\eta_{i j})$. Because of the
  stochastic domination $\mu^{\tmop{RC}}_{\tmmathbf{f}}
  \underset{\tmop{stoch}}{\leqslant} \mu^{\tmop{RC}}_{\tmmathbf{w}}$, the
  conclusion $\mu^{\tmop{RC}}_{\tmmathbf{f}} = \mu^{\tmop{RC}}_{\tmmathbf{w}}$
  follows.
\end{proof}

On the other hand we introduce the magnetization
\begin{equation}
  m_{\tmmathbf{w}} = \mu_{\tmmathbf{w}}^{\tmop{RC}} \left( \frac{1}{1 + (q -
  1) e^{- hS_i^{\tmmathbf{\eta}}}} \right) - \frac{1}{q} \label{mf}
\end{equation}
under the infinite volume measure $\mu_{\tmmathbf{w}}^{\tmop{RC}}$ with wired
boundary condition, for any $h > 0$, where $S_i^{\tmmathbf{\eta}}$ is the
number of sites of the cluster of $\tmmathbf{\eta}$ that contains $i$. We let
$m_{\tmmathbf{f}}$ the same quantity under $\mu_{\tmmathbf{f}}^{\tmop{RC}}$.
We consider also $e_{\tmmathbf{f}}$, the mean energy as in (\ref{ef}) and
$\theta_{\tmmathbf{f}} = \theta$ (see (\ref{theta})) the density of the
percolating cluster under the measure $\mu_{\tmmathbf{f}}^{\tmop{RC}}$, and
call $e_{\tmmathbf{w}}$ and $\theta_{\tmmathbf{w}}$ the corresponding
quantities under $\mu_{\tmmathbf{w}}^{\tmop{RC}}$. We can write
$e_{\tmmathbf{f}}$ and $m_{\tmmathbf{f}}$ as increasing limits and
$e_{\tmmathbf{w}}, m_{\tmmathbf{w}}$ and $\theta_{\tmmathbf{w}}$ as decreasing
limits of continuous, increasing functions of $\beta$. For instance,
\[ e_{\tmmathbf{w}} = \frac{1}{p} \lim_{\Lambda \nearrow \mathbbm{Z}^d}
   \mu_{\Lambda, \tmmathbf{w}}^{\tmop{RC}} (\eta_{i j}) \]
and
\[ \theta_{\tmmathbf{w}} = \lim_{\Delta \nearrow \mathbbm{Z}^d} \lim_{\Lambda
   \nearrow \mathbbm{Z}^d} \mu_{\Lambda, \tmmathbf{w}}^{\tmop{RC}} (0
   \overset{\tmmathbf{\eta}}{\leftrightarrow} \partial \Delta) \]
are decreasing limits while the functions $\beta \mapsto \mu_{\Lambda,
\tmmathbf{w}}^{\tmop{RC}} (\eta_{i j})$ and $\beta \mapsto \mu_{\Lambda,
\tmmathbf{w}}^{\tmop{RC}} (0 \overset{\tmmathbf{\eta}}{\leftrightarrow}
\partial \Delta)$ are continuous, increasing. Hence:

\begin{lemma}
  \label{lem-cont}For any $h \geqslant 0$ and $q \geqslant 1$,
  $e_{\tmmathbf{f}}$ and $m_{\tmmathbf{f}}$ are left-continuous functions of
  $\beta$, while $e_{\tmmathbf{w}}, m_{\tmmathbf{w}}$ and
  $\theta_{\tmmathbf{w}}$ are right-continuous.
\end{lemma}

As a consequence of Lemma \ref{lem-fw} the equalities $e_{\tmmathbf{f}} =
e_{\tmmathbf{w}}$, $m_{\tmmathbf{f}} = m_{\tmmathbf{w}}$ and
$\theta_{\tmmathbf{f}} = \theta_{\tmmathbf{w}}$ hold true for all but
countably many $\beta$. In view of Lemma \ref{lem-cont}, the energy (resp. the
magnetization) is continuous at some $\beta$ if and only if it has the same
value under both $\mu^{\tmop{RC}}_{\tmmathbf{f}}$ and
$\mu^{\tmop{RC}}_{\tmmathbf{w}}$. Hence, at a point of discontinuity it is the
case that $\mu^{\tmop{RC}}_{\tmmathbf{f}} \neq
\mu^{\tmop{RC}}_{\tmmathbf{w}}$. But at such points we cannot have
$\theta_{\tmmathbf{f}} = \theta_{\tmmathbf{w}}$ in view of Theorem
\ref{thm-thetafw} below and Theorem \ref{thm-first-order} follows.

\begin{theorem}
  \label{thm-thetafw}The equality $\theta_{\tmmathbf{f}} =
  \theta_{\tmmathbf{w}}$ implies the uniqueness of random cluster measures --
  in other words, $\mu^{\tmop{RC}}_{\tmmathbf{f}} =
  \mu^{\tmop{RC}}_{\tmmathbf{w}}$ when $\theta_{\tmmathbf{f}} =
  \theta_{\tmmathbf{w}}$.
\end{theorem}

Theorem \ref{thm-thetafw} was proven in {\cite{Gri95}} in the case of $h = 0$
(Theorem 5.2 in {\cite{Gri95}} ; see also Theorem 5.16 in {\cite{Gri}} for the
complete construction). The proof given in {\cite{Gri}} applies verbatim in
the present setting.

The reader might be interested as well in a simpler proof of the fact that
$\theta_{\tmmathbf{w}} = 0$ implies the uniqueness of random cluster measures
(Theorem A.2 in {\cite{ACCN2}}) which shows as well that the Kert\'esz line
remains below the line of discontinuous phase transition.

\subsection{An upper bound on the Kert\'esz line}

Our upper bound is based directly on the conditional probabilities
(\ref{RC-ij}) and (\ref{p}). Since
\[ \inf_{\tmmathbf{\pi}} \mu_{\{i j\}, \tmmathbf{\pi}}^{\tmop{RC}}
   (\tmmathbf{\eta}_{i j} = 1) = \mu_{\{i j\}, \tmmathbf{f}}^{\tmop{RC}}
   (\tmmathbf{\eta}_{i j} = 1) = \tilde{p} \overset{\tmop{def} .}{=}
   \frac{p}{p + (1 - p) w (1)^2 / w (2)}, \]
the measure $\mu_{\tmmathbf{f}}^{\tmop{RC}}$ stochastically dominates
independent bond percolation of parameter $\tilde{p}$, and $\tilde{p} > p_c$
ensures that percolation occurs, i.e. that $\theta > 0$. We recall the
notation $\beta_p = - \ln (1 - p_c)$, which yields
\begin{eqnarray}
  \theta > 0 & \Leftarrow & \tilde{p} > p_c \nonumber\\
  & \Leftrightarrow & \frac{\left( 1 + (q - 1) e^{- h} \right)^2}{1 + (q - 1)
  e^{- 2 h}} < \frac{e^{\beta} - 1}{e^{\beta_p} - 1} \nonumber\\
  & \Leftarrow & \left( 1 + (q - 1) e^{- h} \right)^2 < \frac{e^{\beta} -
  1}{e^{\beta_p} - 1} 
\end{eqnarray}
and the upper bound (\ref{eq-upb-BK}) follows.

\subsection{A lower bound on the Kert\'esz line}

The former method yields here the only information that $h_K (\beta) = +
\infty$ for all $\beta \leqslant \beta_p$. Hence we consider another point of
view : we use a joint measure analogous to $\mu_{\Lambda}^{\tmop{ES}}$
(\ref{ES}) and compare the spins which are not of color $1$ to {\tmem{random
defects}}, which have a vanishing density in the limit $h \rightarrow \infty$.

As we aim at a lower bound that holds for non-integer $q \geqslant 1$, we
consider a modified (monochrome) version of $\mu_{\Lambda}^{\tmop{ES}}$ that
gives only two colors to spin configurations $\tmmathbf{s}$. The color $1$
plays effectively the role of a color in the Potts model, and undergoes the
external field. The color $0$ condensates all $q - 1$ other colors (in the
case of integer $q$). Let
\begin{equation}
  \mu^M_{\Lambda} (\tmmathbf{s}, \tmmathbf{\eta}) = \frac{1}{Z^M_{\Lambda}}
  \omega (\tmmathbf{s}, \tmmathbf{\eta})
\end{equation}
where
\begin{equation}
  \omega (\tmmathbf{s}, \tmmathbf{\eta}) = \prod_{\langle i, j \rangle}^{_{}}
  \left( e^{- \beta} \delta_{\eta_{i j}, 0} + (1 - e^{- \beta})
  \delta_{\eta_{i j}, 1} \delta_{s_i, s_j} \right) \times \prod_i e^{h
  \delta_{s_i, 1} } \times (q - 1)^{N_0 (\tmmathbf{s}, \tmmathbf{\eta})},
\end{equation}
and $N_0 (\tmmathbf{s}, \tmmathbf{\eta})$ is the number of clusters of
$\tmmathbf{\eta}$ that have spin $s = 0$. The marginal law of
$\tmmathbf{\eta}$ equals $\mu_{\Lambda, \tmmathbf{f}}^{\tmop{RC}}$, while the
conditional law of $\tmmathbf{\eta}$ knowing $\tmmathbf{s}$ is the following:
$\tmmathbf{\eta}$ has all edges closed between regions of $\tmmathbf{s}$ of
different colors, while its restriction to the regions with $s = 1$ follows a
bond percolation process of parameter $p = 1 - e^{- \beta}$, and its
restriction to the regions with $s = 0$ follows the usual random cluster
measure of parameters $p = 1 - e^{- \beta}$ and $q' = q - 1$ with free
boundary conditions.

We compare now the structure of spins of color $0$ to independent site
percolation of low density. Let $\bar{\tmmathbf{s}}$ a spin configuration with
$\bar{s}_i = 1$, and call $\tilde{\tmmathbf{s}}$ the modified configuration
with $\tilde{\tmmathbf{s}}_i = 0$. For any $\tmmathbf{\eta}$ such that
$\eta_{i j} = 0$ for all $j$ adjacent to $i$, one has
\begin{equation}
  \omega ( \tilde{\tmmathbf{s}}, \tmmathbf{\eta}) = \omega (
  \bar{\tmmathbf{s}}, \tmmathbf{\eta}) (q - 1) e^{- h} .
\end{equation}
Therefore,
\begin{eqnarray}
  \frac{\mu^M_{\Lambda} ( \tilde{\tmmathbf{s}})}{\mu^M_{\Lambda} (
  \bar{\tmmathbf{s}})} & \geqslant & (q - 1) e^{- h} 
  \frac{\sum_{\tmmathbf{\eta}: \eta_{i j} = 0, \forall j \sim i} \omega (
  \bar{\tmmathbf{s}}, \tmmathbf{\eta})}{\sum_{\tmmathbf{\eta}} \omega (
  \bar{\tmmathbf{s}}, \tmmathbf{\eta})} \nonumber\\
  & = & (q - 1) e^{- h} \times \mu^M_{\Lambda} \left( \eta_{i j} = 0 \text{,
  for all $j$ adjacent to $i$} | \tmmathbf{s}= \bar{\tmmathbf{s}} \right) . 
  \label{muM0}
\end{eqnarray}
But the latter probability is at least $e^{- 2 \beta d}$ and (\ref{muM0})
implies that
\begin{equation}
  \sup_{\bar{\tmmathbf{s}}} \mu \left( s_i = 1| s_j = \bar{s}_j, \forall i
  \neq j \right) \leqslant \rho \overset{\tmop{def} .}{=} \frac{1}{1 + (q - 1)
  e^{- 2 \beta d} e^{- h}} .
\end{equation}
Hence we have a lower bound on the density of defects : the process of good
sites ($s_i = 1$) is stochastically dominated by site percolation of parameter
$\rho$, and percolation cannot occur (i.e. $\theta = 0$) if the mixed
percolation process {\cite{CS}} of site density $\rho$ and edge density $p = 1
- e^{- \beta}$ does not percolate, that is, if there is no infinite cluster
after the removal of closed sites and closed bonds.

The order in which sites and bonds are close does not modify the properties of
the mixed percolation process. Here we shall consider that the edge
percolation at density $p$ is done first, giving the diluted graph $G$ made of
the open edges and their vertices, and that the site percolation of parameter
$\rho$ is realized afterwards. It has been known for a long time that bond
percolation of parameter $\rho$ on $G$ is more likely to succeed than site
percolation (see {\cite{Ham,OW}} for inductive proofs and {\cite{GS}}, proof
of Lemma 5 for a {\tmem{dynamical coupling}}). But the process of bond
percolation with intensity $\rho$ on the diluted graph $G$ boils down to the
classical bond percolation on $\mathbbm{Z}^d$ with parameter $p \times \rho$
and we have shown that
\begin{eqnarray}
  \theta = 0 & \Leftarrow & p \times \rho < p_c \nonumber\\
  & \Leftrightarrow & e^{- \beta_p} - e^{- \beta} < p_c (q - 1) e^{- 2 \beta
  d} e^{- h} 
\end{eqnarray}
which leads to the lower bound (\ref{eq-lwb-BK}).

\bibliographystyle{unsrt}
\bibliography{biblioRW}

\end{document}